\RequirePackage{fix-cm}
\documentclass[twocolumn]{svjour3}          
\smartqed  
\usepackage{graphicx}
\usepackage{natbib}
\begin{document}

\title{Efficiency of size-dependent particle separation by pinched flow fractionation}

\titlerunning{Efficiency of  pinched flow fractionation}        

\author{Aparna Srivastav \and Thomas Podgorski \and Gwennou Coupier}


\institute{
             Laboratoire Interdisciplinaire de Physique\\
             CNRS et Universit\'e J. Fourier - Grenoble I\\
              BP 87, 38402 Saint-Martin d'H\`eres, France \\
              Tel.: +33-476514760\\
              \email{gwennou.coupier@ujf-grenoble.fr}     }

\date{Received: date / Accepted: date}

\maketitle

\begin{abstract}
 Pinched flow fractionation is shown to be an efficient and selective way to quickly separate particles by size in a very polydisperse semi-concentrated suspension. In an effort to optimize the method, we discuss the quantitative influence of the pinching intensity in the balance between the requirements of selectivity and minimal dilution.
\end{abstract}

\section{Introduction}
Sorting micro-particles of different sizes for analytic and preparative purposes has become a great challenge in the fields of chemical or biology research, or even  industrial production. The development of microfluidic techniques has enabled control of flows at a scale similar to the size of the particles together with the use of small amounts of fluid, a target in the process of improving efficiency and reducing costs and dilution of samples.

As reviewed in (\cite{pamme07,kersaudykerhoas08}), several systems were recently developed in that purpose. Generally one expects a quick and precise separation of large quantities of particles, together with low cost and human intervention. In addition, the ability to handle highly concentrated suspensions can also be considered as a requisite. Among these systems, pinched flow fractionation (PFF), which was initially proposed in \cite{yamada04}, has the advantage of being a continuous process based only on hard-core interaction between particles and walls (\cite{luo11}) which is optimized by dedicated flow control. In particular, no use of an external field such as gravity or pressure waves is needed, so that no specific particle property is required.

\begin{figure}[!t]
\centering
\textsl{  \includegraphics[width=\columnwidth]{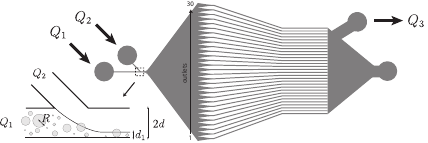}
}  \caption{Scheme of the sorting device. The particles suspension is injected with flow rate $Q_1$ and is pinched by a particle-free fluid with flow rate $Q_2$ in a segment of width $2d$. The suspension then enters a broad segment and splits into 30 outlets. In our experiment, $2d=70\mu$m and the thickness of the channels is $99\mu$m. Drainage with flow rate $Q_3$ was also added, but its effect is not discussed here.}
  \label{fig:schema}
\end{figure}

The principle of PFF is extremely simple (see Figure \ref{fig:schema}): the suspension of (spherical) particles to be sorted is pushed against a wall by a particle-free pinching fluid. Due to their finite size, the centers of the particles are then located on the flow streamline one particle radius distant from the wall.  In principle, collecting each of these streamlines leads to collecting each subpopulation of the sample. In practice, two geometrical improvements are necessary. As most of the pinching fluid will remain particle-free, it is convenient to limit its volume. The pinching is thus realized in a channel (the pinched segment) whose width $2d$ should be similar to the diameter $2R_{\max}$ of the largest particle to be sorted. Secondly, collecting subsamples with narrow size distributions requires that the lateral distance between the different populations is increased.  This is achieved through a broadening of the pinched segment, allowing to add downstream collecting channels (\cite{andersen09}).

As at least the top half of the fluid will remain particle-free, it is convenient to drain it through a single outlet, in order to maximize the space available for the expansion of particle-charged streamlines (\cite{takagi05,vig08,maenaka08}). Alternatively, Sai et al. added microvalves to close some collecting outlets in order to better control the final destination of the particles (\cite{sai06}). At low Reynolds number, which generally applies to such systems, the final destinations in the collecting set-up of the fluid streamlines  created after pinching should theoretically not depend on the geometry of the pinched segment and its expansion. Some works however explored possible geometrical variations and their consequences on separation efficiency as compared to the expected one (\cite{yamada04,zhang06,maenaka08}). Anyhow, it is likely that the small variations observed are mainly due to channel imperfections, 3D effects or optical errors (\cite{jain08}).

In most of the above cited papers, the proposed set-up is validated by injecting a very dilute mixture of a small number  (between 2 and 4) of subpopulations of spherical particles with very different radii. Noting $\lambda$ the ratio between the radii of a large particle and a smaller one, the following papers include such a validation:  \cite{yamada04}, $\lambda=2$; \cite{takagi05}, $\lambda\ge1.4$; \cite{zhang06}, $\lambda=2.5$; \cite{sai06}, $\lambda\ge1.4$; \cite{jain08}, $\lambda=1.5$; \cite{morijiri11}, $\lambda=1.7$. Vig and Kristensen considered seven subpopulations which are mixed and sorted, but their concentration is very low (0.05\%) (\cite{vig08}). As a result, those studies do not fully validate all expected performances of a sorting device, namely the possibility to separate efficiently particles out of a continuum of sizes in a concentrated sample while avoiding excessive dilution in the process. As concentration may lead to increased interactions between particles and possibly decrease the selectivity of the system, the influence of this parameter needs to be explored.

Maenaka \textit{et al.} have considered an emulsion with continuous droplet radius distribution between 0 and 30 $\mu$m, with a concentration of 5\% (\cite{maenaka08}). The sample is injected at a flow rate $Q_1=300 \mu$L/h, with a confining flow $Q_2=9 Q_1$ (see Figure \ref{fig:schema}), and three collecting channels are used. The size distributions in these three channels are quite well separated, although overlappings of around 5 microns seem to occur. As the initial population is only split into three subpopulations, the monodispersity, that is, the existence of a narrow size distribution relatively to the mean size,  remains weak in each of them .

In this paper, we wish to go further and optimize the flow rate ratio $Q_1/Q_2$   between the suspension and the pinching fluid, in order to find the good balance between the requirements of selectivity and minimal dilution. In the meantime, we show that subsamples with good monodispersity can be quickly obtained.

\section{Experimental set-up}

As test particles, we use lipid vesicles obtained through the standard electroformation technique (\cite{angelova92}), which straightforwardly produces polydisperse samples with a continuous size distribution. In order to ensure spherical shapes and prevent lateral migration in the pinched segment due to viscous forces (\cite{coupier08}), a sucrose solution is encapsulated and vesicles are immersed in a slightly hypo-osmotic glucose solution. Being spherical, the vesicles can only deform if their membrane is stretched. The ability of the flow to stretch the membrane is given by the capillary number $C_a=\eta U/\kappa_s$, where $\eta$ is the fluid viscosity, $U$ its typical velocity, and $\kappa$ the membrane stretching modulus, of order $0.3$ N/m for DOPC membranes (\cite{rawicz00}). The maximum flow rate considered here will be 20 mL/h, in a channel of typical size $100\times100$ $\mu$m, therefore $C_a$ will be of order $10^{-3}$ at maximum, and vesicle deformation can be neglected. The use of different fluids inside and outside the vesicles also allows particle visualization through a phase contrast microscope coupled to a fast camera. The sorting device is a standard PDMS microfluidic chip bounded to a glass plate (Figure \ref{fig:schema}). The pinched segment has  width  $2d=70\mu$m and thickness $99\mu$m. From now on, we  set  $d$ as the lengthscale of the problem. The sorted subsamples are observed in 30 collecting channels located at the end of the broadened segment. In this proof of concept device with no sample collection, they all converge to a unique outlet at atmospheric pressure. Additional drainage through a sucking with flow rate $Q_3=0.9  (Q_1+Q_2)$ was added in order to increase the number  of channels with particles. Note that the best location of this drainage is a complex issue as an infinity of geometrical variations are possible (\cite{yamada04,takagi05,vig08,maenaka08}). We shall not explore this in this work.

The particles have radii lying between 0 and 0.97  (in $d$ unit), and split into the 16 first channels (which indicates that our drainage is not optimal). In each outlet channel the mean radius $<R>_i$ of the particles and the standard deviation is measured. The flow rate $Q_2$ was kept constant to $10$ mL/h, and $Q_1$ was varied between $0.2$ and $10$ mL/h. We wish to find the optimum ratio $Q_1/Q_2$ but this parameter  would depend on the chosen width $2d$ (larger $d$ would require larger $Q_2$ for the same pinching efficiency), which depends itself on the maximum size of the particles to be sorted. A more appropriate parameter is therefore the width $d_1$ occupied in the pinched-segment by the fluid coming from channel 1. Roughly,  $d_1$ should be of the order of the radius of the smallest particles to be sorted. 

Finally, initial volume concentrations of 0.8 and 4.8 $\%$ are considered.

\section{Results}

\begin{figure}[!t]
\centering
  \includegraphics[width=\columnwidth]{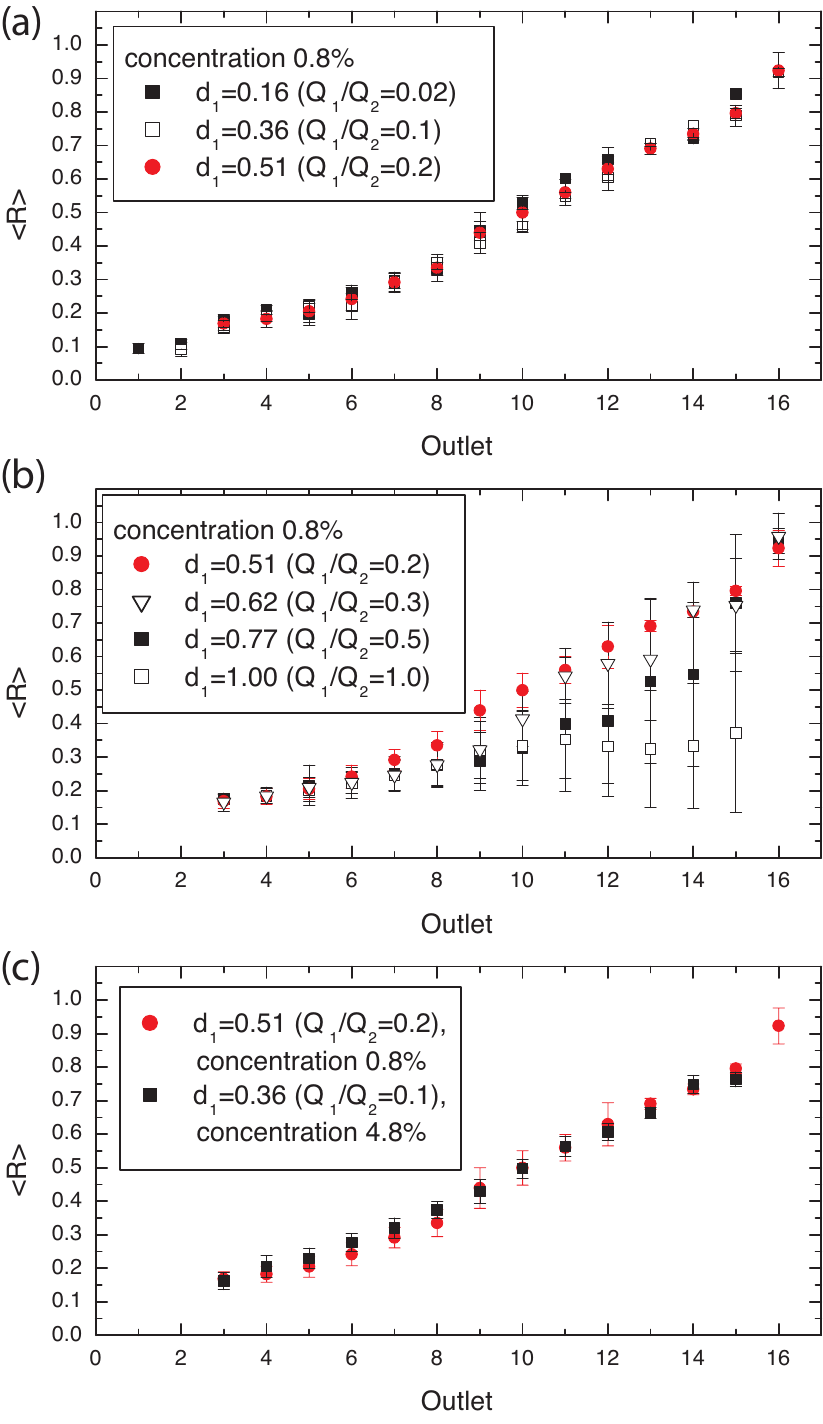}
  \caption{(color online) Size distributions in the outlet channels. (a) and (b): results for initial concentration of 0.8\% and varying pinched width $d_1$ (that is, varying $Q_1/Q_2$). (c): results for  initial concentration of 4.8\% compared with previous ones.}
  \label{fig:res}
\end{figure}

Figures \ref{fig:res}(a) and \ref{fig:res}(b) show the mean sizes of the particles in the different channels for different values of $d_1$ between 0.16 ($Q_1=0.2$ mL/h) and 1 ($Q_1=10$ mL/h). For $d_1\le 0.51$, good separation is achieved, with standard deviations of the order of the half distance between two neighboring mean values. 

In addition to this separation, the obtained subsamples have a good monodispersity: the particles of radii ranging from 0.05 to 0.97 split into 16 subsamples where the mean radius increases quasi linearly  and the standard deviation is roughly constant in the different channels and equal to 0.017 for $d_1=0.16$, $0.021$ for $d_1=0.36$ and to a still reasonable 0.036 for $d_1=0.51$.  Note that if particles of radii between 0 and 1 are perfectly separated into 16 subpopulations, the size difference between the smallest and the largest particle in each subpopulation is $0.063$, which gives a standard deviation of  around $0.016$, so with $d_1=0.16$ we reach the best possible monodispersity with our choice of outlet channels.

The monodispersity quality can be estimated by the ratio between the standard deviation and the mean radius, and from channel 1 to channel 16 we find that it goes  from 17\% to 2\% when $d_1=0.16$ and  from 36\% to 4\% when $d_1=0.51$. The low quality for small particles is related to the choice of outlet channels with equal widths, but the good quality obtained for larger particles show that this is just a scale issue that would be solved by local refinements at the level of the outlets receiving the smallest particles.

\begin{figure}[!t]
\centering
  \includegraphics[width=\columnwidth]{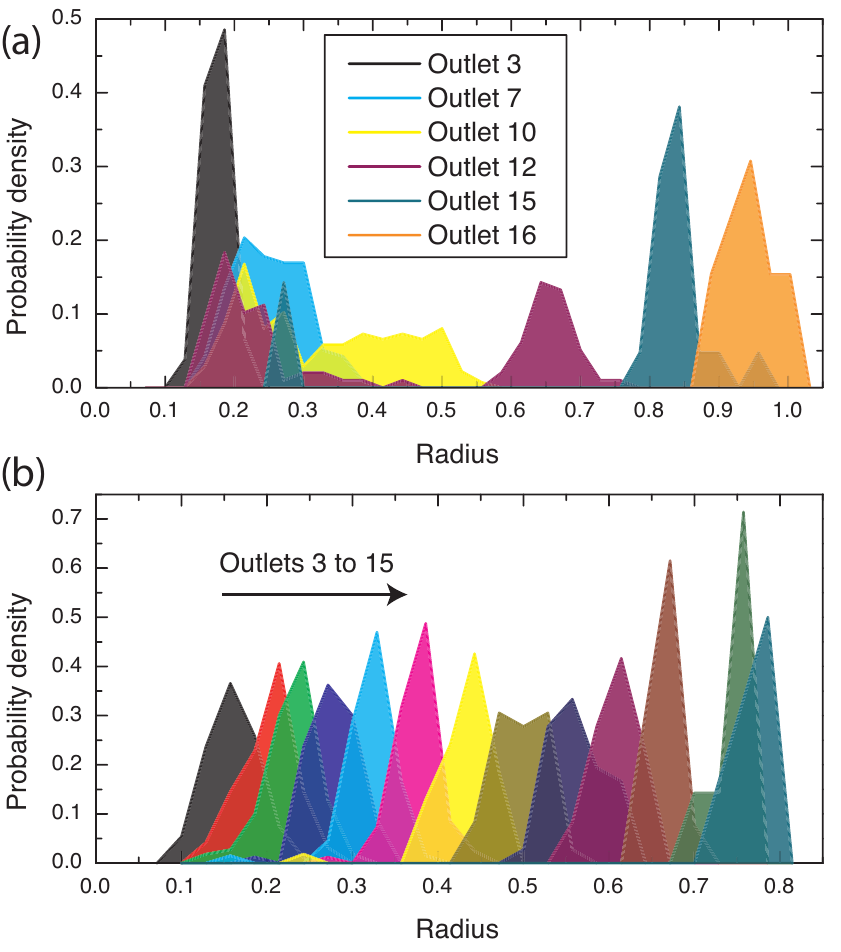}
  \caption{(color online) Particle probability densities $n_i(R)$ in the outlets. $n_i(R)dR$ is the probability to get a vesicle of radius between $R$ and $R+dR$ when picking up one in channel $i$. (a): for some selected outlets for a concentration of 0.8\% and $d_1=0.77$ (same data as in \ref{fig:res}(b)). (b): in the 14 outlets analyzed for a concentration of 4.8\% and $d_1=0.36$ (same data as in \ref{fig:res}(c)). }
  \label{fig:distr}
\end{figure}

As $d_1$ is increased even more, the standard deviations increase drastically while the mean radius varies less between the outlets, in particular for high radii, so separation is not achieved. Note that, for $d_1=0.62$, outlets 14 to 16 have narrow standard deviations, as well as outlet 16 for $d_1=0.77$: the pinching is still strong enough to prevent smaller particles from being at the same level as the large particles that enter these outlets. For intermediate channels (around channel 10) or for $d_1=1$, the population is roughly bidisperse. This can be seen in Figure \ref{fig:distr}(a): the narrow distribution of outlet 3 becomes wider in outlet 7 and 10 and two distinct populations appear in outlet 12 and 15. A possible explanation of this phenomenon can be reached through purely geometrical considerations if one takes into account the effects of walls in inlet 1, before pinching. In this area, the distribution of small particles extends to closer distances to the top wall than larger ones. If the pinching is not strong enough, a reminiscence of this distribution will be found in the pinched fluid, with some small particles higher than larger ones. On the other hand, the bottom wall in the pinched segment prevent the largest particles from respecting this initial order and they must stay in upper position, where non-pinched small particles lie.

Finally, it is remarkable that separation with $d_1$ around 0.5 is almost as good as with the more intuitive choice of $d_1=0.16$, which is of the same order as the smallest particles considered here (even smaller ones, that enter channel 1, were barely distinguishable). If all the small particles present in the initial suspension were present in the whole pinched area of width $d_1=0.51$, they should be present until channel 10, where $<R>=0.50$, which is clearly not the case. One must therefore admit that collective effects take place, so that for instance small particles could be pushed against the wall by larger particles.

Finally, as shown in Figures \ref{fig:res}(c) and  \ref{fig:distr}(b), good separation and monodispersity are also achieved for a concentration of 4.8\%: standard deviations are 0.027 with $d_1=0.36$, which is comparable with the more dilute case. The monodispersity quality reaches 3\% for the larger particles.

\section{Conclusion}

We have shown that pinched flow fractionation is an efficient technique to separate a semi-concentrated polydisperse suspension of micrometric spherical particles into subsamples with tiny overlappings. In addition, monodisperse suspensions with a few \% of variation in sizes can be obtained from that initial suspension where size variations reach 50\%. Moreover, broader initial size distribution would probably not increase the final monodispersity quality in each subsample since the main issue is to separate small to medium sized particles. According to the desired monodispersity quality of the final subsamples the location and width of the collecting outlets can be optimized. Here, we considered outlets of equal width, which resulted in constant standard deviation between subsamples, thus low relative monodispersity quality in the small particles samples. Getting constant monodispersity quality requires to consider outlet of linearly increasing width. If one wishes to get concentrated samples at the outlet, a pinched suspension  width 4 or 5  times larger than the radius of the smallest particles surprisingly appears to be a good compromise: confining more tightens a little bit the distributions, but dilutes more the samples, while a weaker confinement leads to bad sorting. These conclusions are valid for a semi-concentrated suspension (concentrations around 5\%) and for quite high flow rates (some mL/h, that is one order of magnitude higher than in  \cite{maenaka08}). Note that for $Q_1+Q_2=11$  mL/h, the Reynolds number is of order 100 in the pinched segment, so lateral drift of inertial origin could have perturbed the sorting (\cite{segre61}).  However, inertial lift or viscous lift (that can occur in the case of deformable particles) often increase with particle size, therefore they should preserve the sorting effect and only shift the location of collecting outlets.

In addition, we have shown that bidisperse suspensions can be easily obtained through weaker pinching.

\end{document}